\documentclass{cimento}

\usepackage{graphicx}
\usepackage{amsmath,amssymb}
\usepackage{mathrsfs}
\usepackage{bm}
\usepackage{float}
\usepackage{wrapfig}
\usepackage{color}
\title{Radiative corrections to the Higgs couplings in the triplet model\thanks{This proceedings is based on Ref.~\cite{AKKY}}}
\author{M.~Kikuchi,}
\instlist{\inst{}Department of Physics, University of Toyama, 3190 Gofuku, Toyama 930-8555, JAPAN}

\PACSes{
\PACSit{12.15.Lk}{12.60.Fr}
}

\begin{document}

\maketitle

\begin{abstract}
The feature of extended Higgs models can appear in the pattern of deviations 
 from the Standard Model (SM) predictions in coupling constants of the SM-like Higgs boson ($h$). 
We can thus discriminate extended Higgs  models by precisely measuring the pattern of deviations in the coupling constants of $h$,
 even when extra bosons are not found directly.
 In order to compare the theoretical predictions to the future precision data at the ILC,
we must evaluate the theoretical predictions with radiative corrections in various extended Higgs models.  
In this talk, we give our comprehensive study for radiative corrections to various Higgs boson couplings of $h$ in the minimal Higgs triplet model (HTM). 
First, we define renormalization conditions in the model, 
and we calculate the Higgs coupling; 
 $g\gamma\gamma, hWW, hZZ$ and $hhh$ at the one loop level. 
We then evaluate deviations in coupling constants of the SM-like Higgs boson from the predictions in the SM.
We find that one-loop contributions to these couplings are substantial as compared to their expected measurement accuracies at the ILC.
Therefore the HTM has a possibility to be distinguished from the other models by comparing the pattern of deviations in the Higgs boson couplings. 
\end{abstract}

\section{Introduction}
The Higgs boson was discovered at the LHC in July, 2012.
The data indicate that the Higgs boson is the standard model(SM)-like Higgs boson ($h$)~\cite{Higgs_AC}.
However, it does not necessarily mean that the SM is exactly true, because such a Higgs boson can also be predicted in various new physics models.  
There are no theoretical principles for the minimal Higgs sector with one SU(2) doublet scalar field.

Extended Higgs sectors are often introduced in various scenarios of new physics beyond the SM, some of which are motivated to solve the problems such as tiny neutrino masses, dark matter and/or baryon asymmetry of the Universe.
Namely, exploring the shape of the Higgs sector is a key to test the new physics. 
At future collider experiments, the shape of the Higgs sector is expected to be determined in the bottom-up approach. 

In an extended Higgs sector, couplings of the SM-like Higgs boson can deviate from the ones in the SM due to mixing effects and loop effects of additional particles. 
Features of each model appear in the pattern of how all couplings of $h$ deviate from the SM predictions.  
Thus, there are the possibility to discriminate various extended Higgs model by evaluating patterns of deviations in the SM-like Higgs boson couplings in each model.
At the future collider experiment like the International Linear Collider (ILC), couplings of $h$ are expected to be precisely measured typically by $\mathcal{O}(1)\%$.
In order to determine the Higgs sector by comparing with these future precision measurements, we need evaluations for deviations in the SM-like Higgs boson couplings not only by mixing effects but also effects from radiative corrections.


In this talk, we focus on the minimal Higgs triplet model (HTM) as an example of extended Higgs models.
In this model there is a mechanism to generate masses of neutrinos, which is called as the Type-II seesaw mechanism.
 One of the important features in this model is that the
electroweak rho parameter  deviates from the unity at the tree level due to the nonzero vacuum expectation
value (VEV) of the triplet field $v_\Delta$, therefore $v_{\Delta} \ll v(\simeq 246$GeV).
First, we define renormalization conditions for the one-loop corrections.
Then, we calculate various SM-like Higgs couplings at one-loop level; e.g., $h\gamma\gamma, hZZ, hWW$ and $hhh$. 
We evaluate deviations in these coupling constants from the predictions in the SM in the allowed parameter regions from the electroweak precision data and bounds from the perturbative unitarity, taking into account the vacuum stability.
We then discuss the possibility to test the HTM by measuring the pattern of deviations at the ILC.


\section{Higgs triplet model}
The Higgs sector of the HTM is composed of the isospin doublet field $\Phi$ with 
hypercharge $Y=1/2$ and the triplet field $\Delta$ with $Y=1$. 
The detail of the model is shown in Ref.~\cite{typeII}.
The electroweak rho parameter $\rho$ is predicted at the tree level as, 
\begin{eqnletter}
\rho \equiv \frac{m_W^2}{m_Z^2\cos^2\theta_W}=\frac{1+2v_\Delta^2/v_{\phi}^2}{1+4v_\Delta^2/v_{\phi}^2}, \label{rho_triplet}
\end{eqnletter}
where $v_{\phi}$ and $v_{\Delta}$ are the VEVs of the doublet Higgs field and the triplet Higgs field, respectively, which satisfy the relation $v^2 \equiv v_{\phi}^2 + 2v_{\Delta}^2 \simeq (246\ \textrm{GeV})^2$.
Namely, $\rho$ does not satisfy a relation $\rho=1$ at the tree level.
The experimental value of the rho parameter is quite close to the unity; i.e., $\rho^{\textrm{exp}}=1.0004^{+0.0003}_{-0.0004}$~\cite{PDG}.
We note that $v_{\Delta}$ must be smaller than $v_\Phi$. 
In general, $\rho$ deviates from the unity at the tree level in the model whose Higgs sector includes high representation fields more than the doublet representation, as the HTM.
However, it has been found that there are extended Higgs models including high representation fields which satisfy $\rho=1$ at the tree level; 
e.g., the Georgi-Machacek model~\cite{GM} which has a real triplet field in addition to a complex triplet field, and the model with the isospin septet field~\cite{septet}.

The most general form of the Higgs potential is given by 
\begin{eqnletter}
V(\Phi,\Delta)&=&m^2\Phi^\dagger\Phi+M^2\textrm{Tr}(\Delta^\dagger\Delta)+\left[\mu \Phi^Ti\tau_2\Delta^\dagger \Phi+\textrm{h.c.}\right] 
+\lambda_1(\Phi^\dagger\Phi)^2 \notag\\
&+&\lambda_2\left[\textrm{Tr}(\Delta^\dagger\Delta)\right]^2+\lambda_3\textrm{Tr}[(\Delta^\dagger\Delta)^2]
+\lambda_4(\Phi^\dagger\Phi)\textrm{Tr}(\Delta^\dagger\Delta)+\lambda_5\Phi^\dagger\Delta\Delta^\dagger\Phi, \label{pot_htm}
\end{eqnletter}
where $m$ and $M$ are the dimension full real parameters, $\mu$ is the dimension full complex parameter 
which violates the lepton number, and 
$\lambda_1$-$\lambda_5$ are the coupling constants. 
We here assume that $\mu$ is a real parameter.
Seven physical mass eigenstates appear after field mixing, triplet-like Higgs bosons ($H^{\pm\pm}$, $H^\pm$, $A$, $H$) and the SM-like Higgs boson $h$. 

In the case with $v_\Delta \ll v_\Phi$, a characteristic mass hierarchy is realized among the triplet-like Higgs bosons\cite{AKKY,AKKY1} by neglecting $\mathcal{O}(v_\Delta^2/v_\phi^2)$ terms as 
\begin{eqnletter}
m_{H^{++}}^2-m_{H^{+}}^2&=m_{H^{+}}^2-m_{A}^2~~\left(=-\frac{\lambda_5}{4}v^2\right),
\,\,
m_A^2&=m_{H}^2~~(=M_\Delta^2). \label{eq:mass_relation2}
\end{eqnletter}
The mass hierarchy among the triplet-like Higgs bosons depends on the sign of $\lambda_5$.
There are two possibilities for the mass hierarchy, namely the case where $H^{++}$ is the lightest of all triplet-like Higgs bosons ($m_A\, >\, m_{H^+}\, >\, m_{H^{++}}$) and the opposite case ($m_{H^{++}}\, >\, m_{H^+}\, >m_{A}$)\cite{AKKY,AKKY1,KY,lambda5}.
We call the former case (latter case) as Case~I (Case~II). 
We define $\Delta m$ as the mass difference between the singly charged Higgs boson and the lightest triplet-like Higgs boson; i.e., $\Delta m\equiv m_{H^+}-m_{\textrm{lightest}}$.

\section{Renormalization calculations}
In order to calculate 
the renormalized couplings of $hWW, hZZ$ and $hhh$, 
we need to determine counterterms of eight physical parameters and three wave function renormalizations.
We here define on-shell renormalization conditions to determine these counterterms. 
First, we discuss the renormalization of the electroweak sector~\cite{AKKY,KY}. 
Second, we discuss the renormalization of parameters in the Higgs potential~\cite{AKKY,AKKY1}.

\subsection{Electroweak Parameters}
In the case with $\rho_{\textrm{tree}}=1$ like the SM, if we impose renormalization conditions for $m_W, m_Z$ and $\alpha_{\textrm{em}}$, counterterms of other electroweak parameters can be determined by electroweak relations at the tree level~\cite{Hollik-SM}.
For instance, counterterm of the Weinberg angle $\delta (\sin^2\theta_W)$ is determined by using $\rho_{\textrm{tree}}=1$.
On the other hand, in the case with $\rho_{\textrm{tree}} \neq 1$ like the HTM, 
we cannot determine $\delta (\sin^2\theta_W)$ by the same method as in the SM.
However, there is a relation,
\begin{eqnletter}
\cos^2\theta_W = \frac{2m_W^2}{m_Z^2(1+\cos^2\beta')}, \label{swsq_2}
\end{eqnletter}
where $\beta'$ is the mixing angle among CP-odd scalar bosons.
We then use Eq.~(\ref{swsq_2}) to determine $\delta (\sin^2\theta_W)$.
We discuss how the counterterm of $\beta'$ is determined in the renormalization of the Higgs potential.  
This is the difference in the renormalization scheme between the model with $\rho_{\textrm{tree}}=1$ and the HTM.
In addition, 
we can calculate the renormalized $W$ boson mass by these renormalization conditions in Ref.~\cite{AKKY}.
We find that the mass difference $\Delta m$ is constrained by the LEP/SLC electroweak precision data~\cite{PDG} as $0<\Delta m \lesssim 50$ GeV ($0<\Delta m \lesssim 30\text{ GeV}$) for $v_\Delta \lesssim 1$ GeV, 
$40\text{ GeV}\lesssim \Delta m \lesssim 60$ GeV ($30\text{ GeV}\lesssim \Delta \lesssim 50$ GeV) for $v_\Delta=5$ GeV 
and $85\text{ GeV}\lesssim \Delta m\lesssim 100$ GeV ($70\text{ GeV}\lesssim \Delta m\lesssim 85$ GeV) 
for $v_\Delta=10$ GeV.

\subsection{Higgs Potential}
In the Higgs potential, nine parameters ($v,\ \alpha,\ \beta,\ \beta',\  m_{H^{++}},\\ m_{H^+},\ m_A,\ m_H,\ m_h$) exist, where $\alpha\ (\beta)$ is the mixing angle among  CP-even (charged) scalar bosons.
We determine the counterterm of $v$ by the renormalization in the electroweak parameters.
$\delta\beta$ is determined via the relation with $\delta\beta'$.
Other counterterms can be determined by the on-shell renormalizations conditions in the Higgs potential~\cite{AKKY1}. 
The detail of this renormalization scheme is described in the Ref.~\cite{AKKY}.

\section{Higgs couplings at the one-loop level}
In this section, we discuss couplings of $h$ with the gauge bosons 
($\gamma\gamma$, $W^+W^-$ and $ZZ$) and the Higgs selfcoupling $hhh$ at the one-loop level in the favored 
parameter regions by the unitarity bound, the vacuum stability bound and the 
measured $W$ boson mass discussed in the previous sections.
The mass difference $\Delta m$ is constrained from the perturbative unitarity and the vacuum stability because $\Delta m$ is a function of $\lambda_4$ and $\lambda_5$.
The conditions for the vacuum stability bound have been studied in Ref.~\cite{Arhrib}, where 
we require that 
the Higgs potential is bounded from below in any directions.
The unitarity bound has been derived in Ref.~\cite{Aoki-Kanemura} in the Gerogi-Machacek model which contains the HTM. The unitarity bound in the HTM has also been analyzed in Ref.~\cite{Arhrib}.

First, we discuss an one-loop process: $h \rightarrow \gamma\gamma$~\cite{AKKY, KY, h_to_gg}, whose priority was high in the Higgs search at the LHC.
The current experimental data shows that the signal strength for the Higgs to diphoton mode is $1.6\pm0.3$ at the ATLAS~\cite{ATLAS_new} and
$0.8\pm0.3$ at the CMS~\cite{CMS_new}. 
We can directly detect loop effects of the doubly charged Higgs boson $H^{\pm\pm}$ and the singly charged Higgs boson $H^{\pm}$ on the loop via $h\rightarrow \gamma\gamma$ process. 
In particular, 
the contribution from $H^{\pm\pm}$ loop to the $h\to \gamma\gamma$
is quite important as compared to that from the $H^{\pm}$, because the $H^{\pm\pm}$ 
contribution is roughly 4 times larger than that from the $H^{\pm}$ contribution at the amplitude level. 
Then, we calculate the ratio of the event rate for $h\rightarrow \gamma\gamma$ in the HTM to that in the SM; i.e.,
\begin{eqnletter}
 R_{\gamma\gamma} \equiv \frac{\sigma(gg\rightarrow h)_{\textrm{HTM}}\times
                              BR(h\rightarrow \gamma\gamma)_{\textrm{HTM}}}
                            {\sigma(gg\rightarrow h)_{\textrm{SM}}\times
                              BR(h\rightarrow \gamma\gamma)_{\textrm{SM}}},
\end{eqnletter} 
where $\sigma(gg \rightarrow h)_\textrm{model}$ is the cross section of the gluon fusion process, and $BR(h \rightarrow \gamma\gamma)_\textrm{model}$ is
the branching fraction of the process in a model.

\begin{figure}
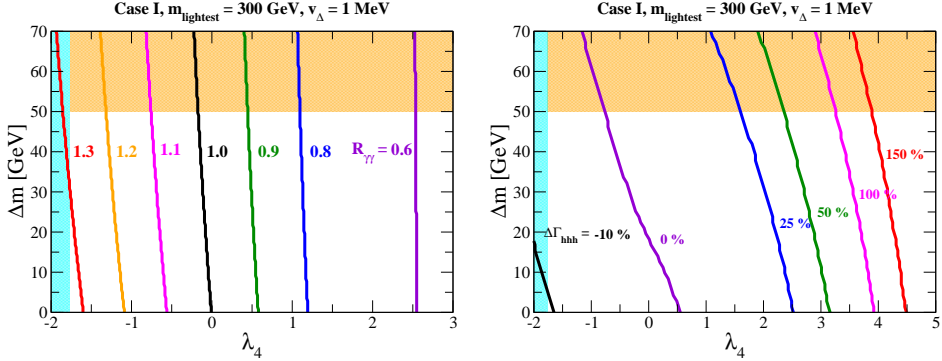

\begin{center}
\includegraphics[width=60mm]{Rgg_m300_vt1mev_1.eps}\hspace{3mm}
\includegraphics[width=60mm]{hhh_ml300_vt1mev_1.eps}
\caption{The left panel (right panel) is contour plots of $R_{\gamma\gamma}$ ($\Delta \Gamma_{hhh}$) for $v_\Delta=1$ MeV and $m_{\text{lightest}}=300$ GeV 
in the $\lambda_4$-$\Delta m$ plane. 
The blue and orange shaded regions are excluded by the vacuum stability bound and the measured $m_W$ data, respectively. 
}
\label{FIG:hgg_hhh}
\end{center}
\end{figure}

In the left panel of Fig.~\ref{FIG:hgg_hhh}, we show the contour plots of $R_{\gamma\gamma}$ for $v_\Delta=1$ MeV and $m_{\text{lightest}}=300$ GeV
on the $\lambda_4$-$\Delta m$ plane in Case I. 
The blue and the orange shaded regions are those excluded by the constraints of the vacuum stability (assuming $\lambda_{2,\ 3}=3$) 
and the experimental data of $m_W$, respectively. 
In this model, $R_{\gamma\gamma}$ is very  sensitive to $\lambda_4$ because the $\lambda_4$ contribution is dominant in parameters of the SM-like  Higgs boson couplings with charged Higgs bosons~\cite{AKKY, h_to_gg}.
The $\Delta m$ dependence in $R_{\gamma\gamma}$ in Case I is small because $m_{H^{++}}$ is fixed.
On the other hand in Case II, $m_{H^{++}}$ become large as $\Delta m$ become large so that $R_{\gamma\gamma}$ slightly depends on $\Delta m$.
We find that the event rate for the $h\rightarrow \gamma\gamma$ process can be several times 10$\%$ larger or smaller than the predictions in the SM. 
Taking into account the CMS data, the parameter region $\lambda_4 \gtrsim -0.5$ is favored in both Case~I and II.

Next, we calculate the Higgs coupling constants at the one-loop level by the renormalization that we discussed at the previous section.
We then define the following quantity to study the deviations of the $hhh$ coupling from the SM predictions: 
\begin{eqnletter}
\Delta \Gamma_{hhh}(p_1^2, p_2^2, q^2)\equiv \frac{\text{Re}\Gamma_{hhh}-\text{Re}\Gamma_{hhh}^{\text{SM}}}{\text{Re}\Gamma_{hhh}^{\text{SM}}}, \label{delhhh}
\end{eqnletter}
where $\Gamma_{hhh}$ is the form factor of the $hhh$ coupling in the HTM, 
$\Gamma_{hhh}^{\textrm{SM}}$ is the corresponding prediction in the SM, $p_1$ and $p_2$ are external incoming momenta, 
and $q$ is the outgoing momentum. 
We fix values of momenta such as $p_1^2=m_h^2,\ p_2^2=m_h^2$ and $q^2=4m_h^2$.
We can define the quantity similar to $\Delta \Gamma_{hhh}$ in the Higgs-Gauge couplings; e.g., $\Delta g_{hVV}$.

The right panel in Fig.~\ref{FIG:hgg_hhh} shows 
the contour plot for 
the deviation of $hhh$ coupling  $\Delta\Gamma_{hhh}$ defined 
in Eq.~(\ref{delhhh}) in Case~I. 
We find that $\Delta \Gamma_{hhh}$ takes positive (negative) values in the case with positive (negative) $\lambda_4$.
$\Gamma_{hhh}$ can deviate $150\%$ from the prediction in the SM  under the constraint from the perturbative unitarity~\cite{Arhrib, Aoki-Kanemura}.
These large deviations in the $hhh$ coupling constant from the non-decoupling property of scalar bosons in the loop, was well known in the two Higgs doublet model~\cite{THDM_reno}. 
Even if we take into account the CMS data of the signal strength for the diphoton mode, $\Delta \Gamma_{hhh}$ can be predicted to be at most about $+50$\%. 
Such a deviation in $\Delta\Gamma_{hhh}$ is expected to be measured at the ILC with a center of mass energy 
to be 500 TeV and integrated luminosity being 500 ab$^{-1}$(ILC500)~\cite{WP}.

The deviations in $hWW$ and $hZZ$ couplings $\Delta g_{hVV} (V=W, Z)$ are predicted to be at most a few percent level in the favored regions by the vacuum stability bound and by the measured $W$ boson mass in Case I and Case II.
 Even if we take into account the LHC data of the signal strength for the diphoton mode, $\Delta g_{hVV}$ can be about 1\%.
The deviations in $hVV$ are expected to be measured at the ILC500~\cite{WP}.

Finally, we discuss the correlation among these SM-like Higgs couplings in the HTM.
We note that contributions to $R_{\gamma\gamma}$ is opposite to the one of $\Delta \Gamma_{hhh}$.
For instance, when $\lambda_4 = -1$ in Case I, 
deviations in the event rate for $h\rightarrow \gamma\gamma$ process is about $+20\%$ and $\Delta \Gamma_{hhh}$ is about $-1\%$.
Furthermore, when $\lambda_4 = 3$ in Case I, 
the event rate for $h\rightarrow \gamma\gamma$ deviates about $-50\%$ and $\Delta \Gamma_{hhh}$ is about $+50\%$.
As shown here, there is the characteristic pattern of deviations in the SM-like Higgs boson couplings in the HTM.
 Namely, this model may be testable by comparing precise theoretical predictions on these coupling constants with precision measurements at future collider experiments, especially at the ILC.

 \section{Conclusions}
We have calculated the decay rate for $h\rightarrow \gamma\gamma$ and the renormalized coupling constants of $hZZ, hWW$ and $hhh$ at the one-loop level in the HTM in order to compare to the data at future collider experiments.
Magnitudes of deviations in these quantities from the predictions in the SM 
have been evaluated in the parameter regions where the unitarity and the vacuum 
stability bounds are satisfied and the predicted $W$ boson mass is consistent with the data.
In the allowed region by the LHC data, deviations in the one-loop corrected $hVV$ and $hhh$ vertices can be about $-1\%$   and $+50\%$, respectively.
We have found features in the HTM by testing the pattern of deviations in coupling constants from the SM predictions.
The HTM has possibilities to be distinguished from the other models by comparing with measuring these deviations in Higgs boson couplings accurately at the ILC.
\\\\

\acknowledgments
I'm very grateful to Shinya Kanemura and Kei Yagyu for the fruitful collaborations of this work.
The work was supported, in part, by Research Fellow of Japan society for the Promotion of Science.

\end{document}